\documentclass[aps,preprint]{revtex4}%
\usepackage{amssymb}
\usepackage{amsfonts}
\usepackage{amsmath}
\usepackage{graphicx}%
\providecommand{\U}[1]{\protect\rule{.1in}{.1in}}
\newtheorem{theorem}{Theorem}

\newenvironment{proof}[1][Proof]{\noindent\textbf{#1.} }{\ \rule{0.5em}{0.5em}}
\begin{document}
\title{The equivalence of Bell's inequality and the Nash inequality in a quantum
game-theoretic setting}
\author{Azhar Iqbal, James M. Chappell, and Derek Abbott}
\affiliation{School of Electrical \& Electronic Engineering, University of Adelaide, South
Australia 5005, Australia.}

\begin{abstract}
The interaction of competing agents is described by classical game theory. It
is now well known that this can be extended to the quantum domain, where
agents obey the rules of quantum mechanics. This is of emerging interest for
exploring quantum foundations, quantum protocols, quantum auctions, quantum
cryptography, and the dynamics of quantum cryptocurrency, for example. In this
paper, we investigate two-player games in which a strategy pair can exist as a
Nash equilibrium when the games obey the rules of quantum mechanics. Using a
generalized Einstein-Podolsky-Rosen (EPR) setting for two-player quantum
games, and considering a particular strategy pair, we identify sets of games
for which the pair can exist as a Nash equilibrium only when Bell's inequality
is violated. We thus determine specific games for which the Nash inequality
becomes equivalent to Bell's inequality for the considered strategy pair.

\end{abstract}
\maketitle

\section{Introduction}

A quantum game \cite{MeyerDavid,EWL,EW,Vaidman,GoogleScholar} describes the
strategic interaction among a set of players sharing quantum states. Players'
strategic choices, or strategies \cite{Binmore,Rasmusen,Osborne}, are local
unitary transformations on the quantum state. The state evolves unitarily and
finally the players' payoffs, or utilities, are obtained by measuring the
entangled state. It turns out that under certain situations sharing of an
entangled quantum state can put the players in an advantageous position and
more efficient outcomes of the game can then emerge. For readers not familiar
with the formalism of quantum theory \cite{Peres}, sharing an entangled state
can be considered equivalent to the situation in which the players have
(shared) access to a `quantum system' having some intrinsically non-classical
aspects. A quantum game would then involve a strategic manoeuvring of the
shared quantum system in which different and perhaps more efficient outcome(s)
of the game can emerge due to non-classical aspect(s) of the shared system.

Now, it is well known that non-classical, and thus apparently strange, aspects
of a shared quantum system can be expressed as constraints on probabilities
relevant to the shared system. Usually expressed as constraints in
correlations, the famous Bell's inequality
\cite{Bell1,Bell2,Bell3,Aspect,CHSH,Peres} can be re-expressed as constraints
on the relevant joint probability and its marginals
\cite{Fine1,Fine2,Halliwell1,Halliwell2}. Essentially, Bell's inequality
emerges as being the necessary and sufficient condition requiring a joint
probability distribution to exist given a set of marginals. It is well known
that Bell's inequality can be violated by a set of quantum mechanical
probabilities---the probabilities that are obtained by the quantum probability
rule. This turns out to be the case even though the quantum probabilities are
normalized as the classical probabilities are. This is because for a set of
marginal (quantum) probabilities that are obtained via the quantum probability
rule, the corresponding joint probability distribution may not exist. The
possibility to express truly non-classical aspects of a quantum system in only
probabilistic terms \cite{Cereceda} has led to suggestions for schemes of
quantum games
\cite{Iqbalepr2,Iqbalepr3,Iqbalepr4,Iqbalepr6,Iqbal,Iqbal1,Iqbal2} that do not
refer to quantum states, unitary transformations, and/or the quantum measurement.

In a classical game allowing mixed strategies, the players' strategies are
convex linear combinations, with real coefficients, of their pure strategies
\cite{Osborne}. Players' strategies in a quantum game \cite{EWL,EW}, however,
are unitary transformations and thus belong to much larger strategy spaces.
This led to the arguments \cite{EnkPike} that quantum games can perhaps be
viewed as extended classical games. In order to obtain an improved comparison
between classical and quantum games, it was suggested \cite{IqbalWeigert} that
the players' strategy sets need to be identical. This has motivated proposals
\cite{Iqbal,Iqbal1,Iqbal2} of quantum games in which players' strategies are
classical, as being convex linear combinations (with real coefficients) of the
classical strategies, and the quantum game emerges from the non-classical
aspects of a shared probabilistic physical system---as expressed by the
constraints on relevant probabilities and their marginals
\cite{Fine1,Fine2,Halliwell1,Halliwell2}.

In the usual approach in the area of quantum games \cite{GoogleScholar}, a
classical game is defined, or given, at the start and its quantum version is
developed afterwards. The usual reasonable requirement being that the
classical mixed-strategy game can be recovered from the quantum game. One then
studies whether the quantum game offers any non-classical outcome(s). In this
paper, the players' strategies in the quantum game remain classical whereas
the new quantum, or non-classical, outcome(s) of the game emerge from the
peculiar quantum probabilities relevant to the quantum system that two players
share to play the game. In contrast to the usual approach in quantum games, in
which the players' strategies are unitary transformations, here we consider a
particular classical strategy pair and then enquire about the set of games for
which that strategy pair can exist as a Nash equilibrium (NE)
\cite{Binmore,Rasmusen,Osborne}. In particular, for a given strategy pair, we
investigate whether there are such games for which that strategy pair can
exist as a NE only when the corresponding Bell's inequality is violated by the
quantum probabilities relevant to the shared quantum system.

We consider two-player games that can be played using the setting of
generalized Einstein-Podolsky-Rosen (EPR) experiments
\cite{Peres,CHSH,Cereceda}. As is known, in this setting a probabilistic
version of Bell's inequality can be obtained
\cite{Fine1,Fine2,Cereceda,Halliwell1,Halliwell2}. We consider particular
strategies and find the sets of games for which the strategies can exist as a
NE only when Bell's inequality is violated. By identifying such games, we show
that there exist strategic outcomes that can only be realized when the game is
played quantum mechanically and also only when the corresponding Bell's
inequality is violated.

The connection between Bell's inequality and the NE was originally reported in
Ref. \cite{Cheon-Iqbal}. However, the Ref. \cite{Cheon-Iqbal} did not use an
EPR setting. In the present paper, we show that the mentioned connection
becomes explicitly direct by using an EPR setting in playing a quantum game.

\section{Two-Player Quantum Games Using the EPR Experiment Setting}

The EPR setting for playing quantum games was introduced in
Ref.~\cite{IqbalWeigert} and was further investigated in
Refs.~\cite{Iqbalepr1,Iqbalepr2,Iqbalepr3,Iqbalepr4,Iqbalepr5,Iqbalepr6,Iqbalepr7,Iqbalepr8}%
. The Refs.~\cite{Iqbalepr2,Iqbalepr3,Iqbalepr4,Iqbalepr6,Iqbal,Iqbal1,Iqbal2}
investigate using the setting of generalized EPR experiments \cite{Cereceda}
for playing quantum games. This setting permits consideration of a
probabilistic version of the corresponding Bell's inequality, which allows
construction of quantum games without referring to the mathematical formalism
of quantum mechanics including Hilbert space, unitary transformations,
entangling operations, and quantum measurements \cite{Peres}. The relationship
between the NE and aspects of Bell's inequality have been indicated in Refs.
\cite{Zhang, Brunner, Pappa}. The present paper's contribution consists of
bringing into focus this relationship and, in particular, finding the specific
games for which this relationship can be explicitly defined. Moreover, in
order to achieve this the present paper uses EPR setting and the probabilistic
version of Bell's inequality.

In the setting of the generalized EPR experiment, Alice and Bob are spatially
separated and are unable to communicate with each other. In an individual run,
both receive one half of a pair of particles originating from a common source.
In the same run of the experiment, both players choose one from two given
(pure) strategies. These strategies are the two directions in space along
which spin or polarization measurements can be made. We denote these
directions to be $S_{1}$, $S_{2}$ for Alice and $S_{1}^{\prime}$,
$S_{2}^{\prime}$ for Bob. Each measurement generates $+1$\ or $-1$\ as the
outcome. Experimental results are recorded for a large number of individual
runs of the experiment. Payoffs are then awarded that depend on the directions
the players choose over many runs (defining the players' strategies), the
matrix of the game they play, and the statistics of the measurement outcomes.
For instance, we denote $\Pr(+1,+1;S_{1},S_{1}^{\prime})$ as the probability
of both Alice and Bob obtaining $+1$ when Alice selects the direction $S_{1}$
whereas Bob selects the direction $S_{1}^{\prime}$. We write $\epsilon_{1}$
for the probability $\Pr(+1,+1;S_{1},S_{1}^{\prime})$ and $\epsilon_{8}$ for
the probability $\Pr(-1,-1;S_{1},S_{2}^{\prime})$ and likewise one can then
write down the relevant probabilities as%

\begin{equation}%
\begin{array}
[c]{c}%
\text{Alice}%
\end{array}%
\begin{array}
[c]{c}%
\underset{}{%
\begin{array}
[c]{c}%
S_{1}%
\end{array}%
\begin{array}
[c]{c}%
+1\\
-1
\end{array}
}\\
\overset{}{%
\begin{array}
[c]{c}%
S_{2}%
\end{array}%
\begin{array}
[c]{c}%
+1\\
-1
\end{array}
}%
\end{array}
\overset{\overset{%
\begin{array}
[c]{c}%
\text{Bob}%
\end{array}
}{%
\begin{array}
[c]{ccc}%
\overset{%
\begin{array}
[c]{c}%
S_{1}^{\prime}%
\end{array}
}{%
\begin{array}
[c]{ccc}%
+1 &  & -1
\end{array}
} &  & \overset{%
\begin{array}
[c]{c}%
S_{2}^{\prime}%
\end{array}
}{%
\begin{array}
[c]{ccc}%
+1 &  & -1
\end{array}
}%
\end{array}
}}{%
\begin{tabular}
[c]{c|c}%
$\underset{}{%
\begin{array}
[c]{cc}%
\epsilon_{1} & \text{ \ \ \ }\epsilon_{2}\\
\epsilon_{3} & \text{ \ \ \ }\epsilon_{4}%
\end{array}
}$ & $\underset{}{%
\begin{array}
[c]{cc}%
\epsilon_{5} & \text{ \ \ \ }\epsilon_{6}\\
\epsilon_{7} & \text{ \ \ \ }\epsilon_{8}%
\end{array}
}$\\\hline
$\overset{}{%
\begin{array}
[c]{cc}%
\epsilon_{9} & \text{ \ \ }\epsilon_{10}\\
\epsilon_{11} & \text{ \ \ }\epsilon_{12}%
\end{array}
}$ & $\overset{}{%
\begin{array}
[c]{cc}%
\epsilon_{13} & \text{ \ \ }\epsilon_{14}\\
\epsilon_{15} & \text{ \ \ }\epsilon_{16}%
\end{array}
}$%
\end{tabular}
\ \ \ }. \label{EPR probabilities}%
\end{equation}
Being normalized, EPR probabilities $\epsilon_{i}$ satisfy the relations%

\begin{equation}
\Sigma_{i=1}^{4}\epsilon_{i}=1,\text{ }\Sigma_{i=5}^{8}\epsilon_{i}=1,\text{
}\Sigma_{i=9}^{12}\epsilon_{i}=1,\text{ }\Sigma_{i=13}^{16}\epsilon_{i}=1.
\label{Normalization}%
\end{equation}
Consider in (\ref{EPR probabilities}), for instance, the case when Alice plays
her strategy $S_{2}$ and Bob plays his strategy $S_{1}^{\prime}$. The two arms
of the Stern-Gerlach apparatus are rotated along these two directions and the
quantum measurement is performed. According to the above table, the
probability that both experimental outcomes are $-1$ is then $\epsilon_{12}$.
Similarly, the probability that the observer $1$'s experimental outcome is
$+1$ and observer $2$'s experimental outcome is $-1$ is given by
$\epsilon_{10}$. The other entries in (\ref{EPR probabilities}) can similarly
be explained. In the present paper, the EPR setting is enforced and that the
players can only choose between two directions.

We now consider a game between two players Alice and Bob, which is defined by
the real numbers $a_{i}$ and $b_{i}$ for $1\leq i\leq16$, and is given by%

\begin{equation}%
\begin{array}
[c]{c}%
\text{Alice}%
\end{array}%
\begin{array}
[c]{c}%
\begin{array}
[c]{c}%
S_{1}%
\end{array}
\\
\\
\\%
\begin{array}
[c]{c}%
S_{2}%
\end{array}
\end{array}
\overset{\overset{%
\begin{array}
[c]{c}%
\text{Bob}%
\end{array}
}{%
\begin{array}
[c]{cccccccccccccccc}%
S_{1}^{\prime} &  &  &  &  &  &  &  &  &  &  &  &  &  &  & S_{2}^{\prime}%
\end{array}
}}{%
\begin{tabular}
[c]{c|c}%
$\underset{}{%
\begin{array}
[c]{ccc}%
(a_{1},b_{1}) &  & (a_{2},b_{2})\\
(a_{3},b_{3}) &  & (a_{4},b_{4})
\end{array}
}$ & $\underset{}{%
\begin{array}
[c]{ccc}%
(a_{5},b_{5}) &  & (a_{6},b_{6})\\
(a_{7},b_{7}) &  & (a_{8},b_{8})
\end{array}
}$\\\hline
$\overset{}{%
\begin{array}
[c]{cc}%
(a_{9},b_{9}) & (a_{10},b_{10})\\
(a_{11},b_{11}) & (a_{12},b_{12})
\end{array}
}$ & $\overset{}{%
\begin{array}
[c]{cc}%
(a_{13},b_{13}) & (a_{14},b_{14})\\
(a_{15},b_{15}) & (a_{16},b_{16})
\end{array}
}$%
\end{tabular}
\ \ \ }. \label{GameDef}%
\end{equation}
For this game, we now define the players' pure strategy payoff relations as%

\begin{align}
\Pi_{A,B}(S_{1},S_{1}^{\prime})  &  =\Sigma_{i=1}^{4}(a_{i},b_{i})\epsilon
_{i},\text{ }\Pi_{A,B}(S_{1},S_{2}^{\prime})=\Sigma_{i=5}^{8}(a_{i}%
,b_{i})\epsilon_{i},\nonumber\\
\Pi_{A,B}(S_{2},S_{1}^{\prime})  &  =\Sigma_{i=9}^{12}(a_{i},b_{i}%
)\epsilon_{i},\text{ }\Pi_{A,B}(S_{2},S_{2}^{\prime})=\Sigma_{i=13}^{16}%
(a_{i},b_{i})\epsilon_{i}, \label{PayoffParts}%
\end{align}
where $\Pi_{A}(S_{1},S_{2}^{\prime})$, for example, is Alice's payoff when she
plays $S_{1}$ and Bob plays $S_{2}^{\prime}$.

It can be seen that in the way it is defined, the game is inherently
probabilistic. That is, in (\ref{GameDef}) the players' payoffs even for their
pure strategies assume an underlying probability distribution as given by
(\ref{EPR probabilities}). Now, we can also define a mixed-strategy version of
this game as follows. Consider Alice playing the strategy $S_{1}$ with
probability $p$ and the strategy $S_{2}$ with probability $(1-p)$ whereas Bob
playing the strategy $S_{1}^{\prime}$ with probability $q$ and the strategy
$S_{2}^{\prime}$ with probability $(1-q)$. Using (\ref{GameDef},
\ref{PayoffParts}) the players' mixed strategy payoff relations can then be
obtained as%

\begin{equation}
\Pi_{A,B}(p,q)=%
\begin{pmatrix}
p\\
1-p
\end{pmatrix}
^{T}%
\begin{pmatrix}
\Pi_{A,B}(S_{1},S_{1}^{\prime}) & \Pi_{A,B}(S_{1},S_{2}^{\prime})\\
\Pi_{A,B}(S_{2},S_{1}^{\prime}) & \Pi_{A,B}(S_{2},S_{2}^{\prime})
\end{pmatrix}%
\begin{pmatrix}
q\\
1-q
\end{pmatrix}
.
\end{equation}
Assuming that the strategy pair $(p^{\star},q^{\star})$ is a NE, we then require%

\begin{equation}
\Pi_{A}(p^{\star},q^{\star})-\Pi_{A}(p,q^{\star})\geqslant0,\text{ \ \ }%
\Pi_{B}(p^{\star},q^{\star})-\Pi_{B}(p^{\star},q)\geqslant0, \label{NE}%
\end{equation}
that takes the form%

\begin{align}
\Pi_{A}(p^{\star},q^{\star})-\Pi_{A}(p,q^{\star})  &  =(p^{\star}-p)%
\begin{pmatrix}
1\\
-1
\end{pmatrix}
^{T}%
\begin{pmatrix}
\Pi_{A}(S_{1},S_{1}^{\prime}) & \Pi_{A}(S_{1},S_{2}^{\prime})\\
\Pi_{A}(S_{2},S_{1}^{\prime}) & \Pi_{A}(S_{2},S_{2}^{\prime})
\end{pmatrix}%
\begin{pmatrix}
q^{\star}\\
1-q^{\star}%
\end{pmatrix}
\geq0,\nonumber\\
\Pi_{B}(p^{\star},q^{\star})-\Pi_{B}(p^{\star},q)  &  =%
\begin{pmatrix}
p^{\star}\\
1-p^{\star}%
\end{pmatrix}
^{T}%
\begin{pmatrix}
\Pi_{B}(S_{1},S_{1}^{\prime}) & \Pi_{B}(S_{1},S_{2}^{\prime})\\
\Pi_{B}(S_{2},S_{1}^{\prime}) & \Pi_{B}(S_{2},S_{2}^{\prime})
\end{pmatrix}%
\begin{pmatrix}
1\\
-1
\end{pmatrix}
(q^{\star}-q)\geq0. \label{NashInequalities}%
\end{align}

Note that the players' strategies are classical whereas the game itself is not
classical as the underlying probabilities of the game are quantum mechanical
as obtained from the EPR experiments. Players' payoffs are defined in terms of
EPR quantum probabilities that can violate Bell's inequalities. Thus a
classical game can in no way model this quantum game. This setting also
circumvents the criticism of Enk and Pike \cite{EnkPike} on quantum games. Enk
and Pike noted that as the players in the quantum game have access to much
larger strategy sets, the quantum game can be considered as another classical
game with an extended set of classical strategies. In the considered setting,
players' strategy sets are identical in both the classical and quantum games
and players' payoff relations are obtained from an underlying probability
distribution that is quantum mechanical.

Although this game is played using the setting of generalized EPR experiments,
in which the players strategies consist of choosing between two directions,
one can notice that under appropriate conditions, the game can be reduced to
the usual mixed-strategy version of the standard two-player two-strategy
noncooperative game. Non-cooperative games \cite{Binmore,Rasmusen,Osborne}
were investigated in the early work \cite{EWL,EW} on quantum games. To see
this, we consider the case when%

\begin{gather}
a_{i}(1\leq i\leq4)=\alpha,\text{ }b_{i}(1\leq i\leq4)=\alpha,\nonumber\\
a_{i}(5\leq i\leq8)=\beta,\text{ }b_{i}(5\leq i\leq8)=\gamma,\nonumber\\
a_{i}(9\leq i\leq12)=\gamma,\text{ }b_{i}(9\leq i\leq12)=\beta,\nonumber\\
a_{i}(13\leq i\leq16)=\delta,\text{ }b_{i}(13\leq i\leq16)=\delta,
\end{gather}
and then from Eqs.~(\ref{PayoffParts}) and (\ref{Normalization}) we obtain%

\begin{gather}
\Pi_{A,B}(S_{1},S_{1}^{\prime})=\alpha\Sigma_{i=1}^{4}\epsilon_{i}%
=\alpha,\nonumber\\
\Pi_{A}(S_{1},S_{2}^{\prime})=\beta\Sigma_{i=5}^{8}\epsilon_{i},=\beta,\text{
}\Pi_{B}(S_{1},S_{2}^{\prime})=\gamma\Sigma_{i=5}^{8}\epsilon_{i}%
,=\gamma,\nonumber\\
\Pi_{A}(S_{2},S_{1}^{\prime})=\gamma\Sigma_{i=9}^{12}\epsilon_{i}%
=\gamma,\text{ }\Pi_{B}(S_{2},S_{1}^{\prime})=\beta\Sigma_{i=9}^{12}%
\epsilon_{i}=\beta,\nonumber\\
\Pi_{A,B}(S_{2},S_{2}^{\prime})=\delta\Sigma_{i=13}^{16}\epsilon_{i}=\delta.
\end{gather}
In view of (\ref{NE}), Nash inequalities for the strategy pair $(p^{\star
},q^{\star})$ then take the form%

\begin{align}
\Pi_{A}(p^{\star},q^{\star})-\Pi_{A}(p,q^{\star})  &  =(p^{\star}-p)%
\begin{pmatrix}
1\\
-1
\end{pmatrix}
^{T}%
\begin{pmatrix}
\alpha & \beta\\
\gamma & \delta
\end{pmatrix}%
\begin{pmatrix}
q^{\star}\\
1-q^{\star}%
\end{pmatrix}
\geq0,\nonumber\\
\Pi_{B}(p^{\star},q^{\star})-\Pi_{B}(p^{\star},q)  &  =%
\begin{pmatrix}
p^{\star}\\
1-p^{\star}%
\end{pmatrix}
^{T}%
\begin{pmatrix}
\alpha & \gamma\\
\beta & \delta
\end{pmatrix}%
\begin{pmatrix}
1\\
-1
\end{pmatrix}
(q^{\star}-q)\geq0,
\end{align}
which give us Nash inequalities for the mixed strategy $(p^{\star},q^{\star})$
for the following symmetric game%

\begin{equation}
\left(
\begin{array}
[c]{cc}%
(\alpha,\alpha) & (\beta,\gamma)\\
(\gamma,\beta) & (\delta,\delta)
\end{array}
\right)  .
\end{equation}
When $\beta<\delta<\alpha<\gamma$ this game results in the well known game of
Prisoners' Dilemma. As is well known, for this game $(p^{\star},q^{\star
})=(0,0)$ comes out as the unique NE.

\subsection{Cereceda's analysis and the probabilitsic version of CHSH sum of
correlations}

A convenient solution of the system (\ref{Normalization},
\ref{LocalityConstraint}) was reported by Cereceda in \cite{Cereceda} and
given in the Appendix A. Cereceda expressed the set of probabilities
$\upsilon=\left\{  \epsilon_{2},\text{ }\epsilon_{3},\text{ }\epsilon
_{6},\text{ }\epsilon_{7},\text{ }\epsilon_{10},\text{ }\epsilon_{11},\text{
}\epsilon_{13},\text{ }\epsilon_{16}\right\}  $ in terms of the remaining set
of probabilities i.e.%

\begin{equation}
\mu=\left\{  \epsilon_{1},\text{ }\epsilon_{4},\text{ }\epsilon_{5},\text{
}\epsilon_{8},\text{ }\epsilon_{9},\text{ }\epsilon_{12},\text{ }\epsilon
_{14},\text{ }\epsilon_{15}\right\}  , \label{dependentProbabilities}%
\end{equation}
and thus the elements of the set $\mu$ can be considered as independent variables.

In a particular run of the EPR experiment, the requirements of locality
dictate that the outcome of $+1$ or $-1$ (obtained along the direction $S_{1}$
or direction $S_{2}$) is independent of whether the direction $S_{1}^{\prime}$
or the direction $S_{2}^{\prime}$ is chosen in that run. Similarly, the
outcome of $+1$ or $-1$ (obtained along $S_{1}^{\prime}$ or $S_{2}^{\prime}$)
is independent of whether the direction $S_{1}$ or the direction $S_{2}$ is
chosen in that run. These locality requirements when translated in terms of
the probability set $\epsilon_{j}$ can be expressed as Eqs.
(\ref{LocalityConstraint}) in Appendix A.

Relevant to the EPR setting is the Clauser-Horne-Shimony-Holt (CHSH) form of
Bell's inequality that is usually expressed in terms of the correlations
$\left\langle S_{1}S_{1}^{\prime}\right\rangle $, $\left\langle S_{1}%
S_{2}^{\prime}\right\rangle $, $\left\langle S_{2}S_{1}^{\prime}\right\rangle
$, $\left\langle S_{2}S_{2}^{\prime}\right\rangle $. Using
(\ref{EPR probabilities}) the correlation $\left\langle S_{1}S_{1}^{\prime
}\right\rangle $, for instance, can be obtained as%

\begin{gather}
\left\langle S_{1}S_{1}^{\prime}\right\rangle =\Pr(S_{1}=1,S_{1}^{\prime
}=1)-\Pr(S_{1}=1,S_{1}^{\prime}=-1)\nonumber\\
-\Pr(S_{1}=-1,S_{1}^{\prime}=+1)+\Pr(S_{1}=-1,S_{1}^{\prime}=-1)\nonumber\\
=\epsilon_{1}-\epsilon_{2}-\epsilon_{3}+\epsilon_{4}.
\end{gather}
Expressions for the correlations $\left\langle S_{1}S_{2}^{\prime
}\right\rangle $, $\left\langle S_{2}S_{1}^{\prime}\right\rangle $, and
$\left\langle S_{2}S_{2}^{\prime}\right\rangle $ can similarly be obtained.
The CHSH sum of correlations is given as%

\begin{equation}
\Delta=\left\langle S_{1}S_{1}^{\prime}\right\rangle +\left\langle S_{1}%
S_{2}^{\prime}\right\rangle +\left\langle S_{2}S_{1}^{\prime}\right\rangle
-\left\langle S_{2}S_{2}^{\prime}\right\rangle , \label{CHSH(a)}%
\end{equation}
and the CHSH inequality stating that $\left\vert \Delta\right\vert \leq2$
holds for any theory of local hidden variables.

The set of constraints on probabilities $\epsilon_{i}$ that are imposed by
Tsirelson's bound\emph{\ }\cite{Tsirelson}\emph{\ }state that the quantum
prediction of the CHSH sum of correlations $\Delta$, defined in (\ref{CHSH(a)}%
), is bounded in absolute value by $2\sqrt{2}$ i.e. $\left\vert \Delta
_{QM}\right\vert \leq2\sqrt{2}$. Taking into account \cite{Cereceda} the
normalization condition (\ref{Normalization}), the quantity $\Delta$ can
equivalently be expressed as%

\begin{equation}
\Delta=2(\epsilon_{1}+\epsilon_{4}+\epsilon_{5}+\epsilon_{8}+\epsilon
_{9}+\epsilon_{12}+\epsilon_{14}+\epsilon_{15}-{\small 2}). \label{delta}%
\end{equation}
Bell's inequality can thus be written as $0\leq(2-\left\vert \Delta\right\vert
)$ and is violated when the discriminant $(2-\left\vert \Delta\right\vert
)<0$. Bell's inequality is thus violated when the discriminant attains a
negative value that occurs when either $\Delta>2$ or $\Delta<-2$.

\section{Games for which Nash inequalities involve CHSH sum of correlations}

We note from (\ref{NashInequalities}) that Nash inequalities for the strategy
pair $(p^{\star},q^{\star})=(1/2,1/2)$ take the form%

\begin{gather}
\Pi_{A}(1/2,1/2)-\Pi_{A}(p,1/2)=\nonumber\\
(1/2)(1/2-p)\left[  \Pi_{A}(S_{1},S_{1}^{\prime})+\Pi_{A}(S_{1},S_{2}^{\prime
})-\Pi_{A}(S_{2},S_{1}^{\prime})-\Pi_{A}(S_{2},S_{2}^{\prime})\right]
\geq0,\label{NE1}\\
\Pi_{B}(1/2,1/2)-\Pi_{B}(1/2,q)=\nonumber\\
(1/2)(1/2-q)\left[  \Pi_{B}(S_{1},S_{1}^{\prime})-\Pi_{B}(S_{1},S_{2}^{\prime
})+\Pi_{B}(S_{2},S_{1}^{\prime})-\Pi_{B}(S_{2},S_{2}^{\prime})\right]  \geq0,
\label{NE2}%
\end{gather}
which hold in order for the strategy pair $(p^{\star},q^{\star})=(1/2,1/2)$ to
exist as a NE. Now, the presence of the terms $(1/2-p)$ in (\ref{NE1}) and
$(1/2-q)$ in (\ref{NE2}) forces both expressions within the square brackets to
be identically zero. Nash inequalities for the strategy pair $(p^{\star
},q^{\star})=(1/2,1/2),$ therefore, cannot be expressed in terms of the
discriminant $(2-\left\vert \Delta\right\vert )$. That is, the strategy pair
$(p^{\star},q^{\star})=(1/2,1/2)$ cannot exist as a NE when Bell's inequality
is violated. We therefore consider a second example of the strategy pair
$(p^{\star},q^{\star})=(1,1/2)$ that allows us to establish a direct
connection between Bell's inequality and Nash inequality.

\begin{theorem}
For the set of games for which%
\begin{gather}
\Pi_{A}(S_{2},S_{1}^{\prime})+\Pi_{A}(S_{2},S_{2}^{\prime})-\Pi_{A}%
(S_{1},S_{1}^{\prime})-\Pi_{A}(S_{1},S_{2}^{\prime})=2-\left\vert
\Delta\right\vert ,\label{Games1}\\
\Pi_{B}(S_{1},S_{1}^{\prime})-\Pi_{B}(S_{1},S_{2}^{\prime})=0, \label{Games2}%
\end{gather}
the strategy pair $(p^{\star},q^{\star})=(1,1/2)$ exists as a NE when Bell's
inequality is violated.
\end{theorem}

\begin{proof}
Nash inequalities (\ref{NashInequalities}) for the strategy pair $(p^{\star
},q^{\star})=(1,1/2)$ take the form:%

\begin{gather}
\Pi_{A}(1,1/2)-\Pi_{A}(p,1/2)=\nonumber\\
-(1/2)(1-p)\left[  \Pi_{A}(S_{2},S_{1}^{\prime})+\Pi_{A}(S_{2},S_{2}^{\prime
})-\Pi_{A}(S_{1},S_{1}^{\prime})-\Pi_{A}(S_{1},S_{2}^{\prime})\right]
\geq0,\label{Nash1}\\
\Pi_{B}(1,1/2)-\Pi_{A}(1,q)=\nonumber\\
\left[  \Pi_{B}(S_{1},S_{1}^{\prime})-\Pi_{B}(S_{1},S_{2}^{\prime})\right]
(1/2-q)\geq0. \label{Nash2}%
\end{gather}

\end{proof}

Now, the inequality (\ref{Nash1}) can hold when the term in square bracket is
negative or zero. As Bell's inequality is violated when the discriminant
$(2-\left\vert \Delta\right\vert )$ is negative, therefore, for the set of
games that are defined by the conditions (\ref{Game1}, \ref{Game2}) the
strategy pair $(p^{\star},q^{\star})=(1,1/2)$ exists as a NE when Bell's
inequality is violated.

For the set of games defined by the following conditions the strategy pair
$(p^{\star},q^{\star})=(1,1/2)$ exists as a NE when Bell's inequality is
violated, for $0\leq\Delta,$%
\begin{align}
b_{1}  &  =b_{2}=b_{5}=b_{6}\text{ and }b_{3}=b_{4}=b_{7}=b_{8},\nonumber\\
a_{2}  &  =-a_{5}+a_{12}+a_{15},\text{ \ \ }a_{3}=a_{1}+a_{4}+a_{5}%
-a_{12}-a_{15}-4,\nonumber\\
a_{6}  &  =a_{4}+a_{5}+a_{8}-a_{12}-a_{15}-4,\text{ \ \ }a_{7}=-a_{4}%
+a_{12}+a_{15},\nonumber\\
a_{9}  &  =a_{1}+a_{4}+a_{5}+a_{8}-a_{12}-a_{14}-a_{15}-4,\nonumber\\
a_{10}  &  =a_{4}+a_{8}-a_{14},\text{ \ \ }a_{11}=a_{1}+a_{5}-a_{15}%
,\nonumber\\
a_{13}  &  =-a_{4}-a_{8}+a_{12}+a_{14}+a_{15}+4,\text{ \ \ }a_{16}=a_{4}%
+a_{8}-a_{12}, \label{ConditionsOna1}%
\end{align}
and for $\Delta<0,$ the same is true for the set of games that is defined by
these conditions:%
\begin{align}
b_{1}  &  =b_{2}=b_{5}=b_{6}\text{ and }b_{3}=b_{4}=b_{7}=b_{8},\nonumber\\
a_{2}  &  =-a_{5}+a_{12}+a_{15}-4,\text{ \ \ }a_{3}=a_{1}+a_{4}+a_{5}%
-a_{12}-a_{15}+8,\nonumber\\
a_{6}  &  =a_{4}+a_{5}+a_{8}-a_{12}-a_{15}+8,\text{ \ \ }a_{7}=-a_{4}%
+a_{12}+a_{15}-4,\nonumber\\
a_{9}  &  =a_{1}+a_{4}+a_{5}+a_{8}-a_{12}-a_{14}-a_{15}+12,\nonumber\\
a_{10}  &  =a_{4}+a_{8}-a_{14}+4,\text{ \ \ }a_{11}=a_{1}+a_{5}-a_{15}%
+4,\nonumber\\
a_{13}  &  =-a_{4}-a_{8}+a_{12}+a_{14}+a_{15}-8,\text{ \ \ }a_{16}=a_{4}%
+a_{8}-a_{12}+4. \label{ConditionsOna2}%
\end{align}

\begin{proof}
Using Eqs.~(\ref{PayoffParts}) we write Eqs.~(\ref{Games1}, \ref{Games2}) as,%

\begin{gather}
\Sigma_{i=9}^{12}a_{i}\epsilon_{i}+\Sigma_{i=13}^{16}a_{i}\epsilon_{i}%
-\Sigma_{i=1}^{4}a_{i}\epsilon_{i}-\Sigma_{i=5}^{8}a_{i}\epsilon
_{i}=2-\left\vert \Delta\right\vert ,\label{Games3}\\
\Sigma_{i=1}^{4}b_{i}\epsilon_{i}-\Sigma_{i=5}^{8}b_{i}\epsilon_{i}=0.
\label{Games4}%
\end{gather}
Now, using Eq.~(\ref{dependentProbabilities}), the left sides of
Eq.~(\ref{Games3}) can then be expressed in terms of the probabilities from
the set $\mu$ as follows:%

\begin{gather}
\Sigma_{i=9}^{12}a_{i}\epsilon_{i}+\Sigma_{i=13}^{16}a_{i}\epsilon_{i}%
-\Sigma_{i=1}^{4}a_{i}\epsilon_{i}-\Sigma_{i=5}^{8}a_{i}\epsilon
_{i}=\nonumber\\
(\epsilon_{1}/2)(-2a_{1}+a_{2}+a_{3}-a_{6}+a_{7}-a_{10}+a_{11}-a_{13}%
+a_{16})+\nonumber\\
(\epsilon_{4}/2)(-2a_{4}+a_{2}+a_{3}+a_{6}-a_{7}+a_{10}-a_{11}+a_{13}%
-a_{16})+\nonumber\\
(\epsilon_{5}/2)(-2a_{5}-a_{2}+a_{3}+a_{6}+a_{7}+a_{10}-a_{11}+a_{13}%
-a_{16})+\nonumber\\
(\epsilon_{8}/2)(-2a_{8}+a_{2}-a_{3}+a_{6}+a_{7}-a_{10}+a_{11}-a_{13}%
+a_{16})+\nonumber\\
(\epsilon_{9}/2)(2a_{9}+a_{2}-a_{3}+a_{6}-a_{7}-a_{10}-a_{11}+a_{13}%
-a_{16})+\nonumber\\
(\epsilon_{12}/2)(2a_{12}-a_{2}+a_{3}-a_{6}+a_{7}-a_{10}-a_{11}-a_{13}%
+a_{16})+\nonumber\\
(\epsilon_{14}/2)(2a_{14}-a_{2}+a_{3}-a_{6}+a_{7}+a_{10}-a_{11}-a_{13}%
-a_{16})+\nonumber\\
(\epsilon_{15}/2)(2a_{15}+a_{2}-a_{3}+a_{6}-a_{7}-a_{10}+a_{11}-a_{13}%
-a_{16})+\nonumber\\
(1/2)(-a_{2}-a_{3}-a_{6}-a_{7}+a_{10}+a_{11}+a_{13}+a_{16})=2-\left\vert
\Delta\right\vert . \label{Collect1}%
\end{gather}
Similarly, the left side of Eq.~(\ref{Games4}) now takes the form:%

\begin{gather}
\Sigma_{i=1}^{4}b_{i}\epsilon_{i}-\Sigma_{i=5}^{8}b_{i}\epsilon_{i}%
=\nonumber\\
(\epsilon_{1}/2)(2b_{1}-b_{2}-b_{3}-b_{6}+b_{7})+(\epsilon_{4}/2)(-b_{2}%
-b_{3}+2b_{4}+b_{6}-b_{7})+\nonumber\\
(\epsilon_{5}/2)(b_{2}-b_{3}-2b_{5}+b_{6}+b_{7})+(\epsilon_{8}/2)(-b_{2}%
+b_{3}+b_{6}+b_{7}-2b_{8})+\nonumber\\
(\epsilon_{9}/2)(-b_{2}+b_{3}+b_{6}-b_{7})+(\epsilon_{12}/2)(b_{2}-b_{3}%
-b_{6}+b_{7})+\nonumber\\
(\epsilon_{14}/2)(b_{2}-b_{3}-b_{6}+b_{7})+(\epsilon_{15}/2)(-b_{2}%
+b_{3}+b_{6}-b_{7})+\nonumber\\
(1/2)(b_{2}+b_{3}-b_{6}-b_{7})=0. \label{Collect2}%
\end{gather}
As the probabilities in the set $\mu$ are considered independent, comparing
the two sides of Eq.~(\ref{Collect2}) leads us to obtain:%

\begin{equation}
b_{1}=b_{2}=b_{5}=b_{6}\text{ \textrm{and }}b_{3}=b_{4}=b_{7}=b_{8}.
\label{ConditionsOnb}%
\end{equation}
Consider now the right side of Eq.~(\ref{Collect1}). As $\Delta=2(\epsilon
_{1}+\epsilon_{4}+\epsilon_{5}+\epsilon_{8}+\epsilon_{9}+\epsilon
_{12}+\epsilon_{14}+\epsilon_{15}-{\small 2})$, the discriminant
$(2-\left\vert \Delta\right\vert )$ can be negative for the following two
cases: a) For case $0\leq\Delta,$ we have $2-\left\vert \Delta\right\vert
=2-\Delta$. The rank of the system (\ref{Collect1}) is $7$. We take
$a_{1},a_{4},a_{5},a_{8},a_{12},a_{14},a_{15}$ as independently chosen
constants and compare the coefficients of the independent probabilities in the
set $\mu$ on the two sides of Eq.~(\ref{Collect1}). This gives the set of
conditions (\ref{ConditionsOna1}). That is, for the set of games defined by
the conditions (\ref{ConditionsOnb}, \ref{ConditionsOna1}) the strategy pair
$(p^{\star},q^{\star})=(1,1/2)$ exists as a NE when the Bell's inequality is
violated. For case $\Delta<0,$ we have $\left\vert \Delta\right\vert =-\Delta$
and $2-\left\vert \Delta\right\vert =2+\Delta$. Following the steps from the
last case, we obtain the set of conditions (\ref{ConditionsOna2}). As before,
for the set of games that are defined by the conditions (\ref{ConditionsOnb},
\ref{ConditionsOna2}), the strategy pair $(p^{\star},q^{\star})=(1,1/2)$
exists as a NE when the Bell's inequality is violated.
\end{proof}

\section{An example}

As a specific example, and in view of Eqs.~(\ref{ConditionsOnb}), we assign
the value of $1$ arbitrarily to $b_{1},$ $b_{2},$ $b_{5},$ $b_{6}$ and also
the same value to $b_{3},$ $b_{4},$ $b_{7},$ $b_{8}$. Also, as
Eq.~(\ref{Games4}) does not involve the constants $b_{9},$ $b_{10},$ $b_{11},$
$b_{12},$ $b_{13},$ $b_{14},$ $b_{15},$ $b_{16}$ we also assign the value of
$1$ to them. Likewise, we assign the value of $1$ to the independently chosen
constants $a_{1},$ $a_{4},$ $a_{5},$ $a_{8},$ $a_{12},$ $a_{14},$ $a_{15}$.
With reference to Eqs.~(\ref{ConditionsOna1}, \ref{ConditionsOna2}) we obtain
the following two games,%

\begin{equation}%
\begin{array}
[c]{c}%
\text{Alice}%
\end{array}%
\begin{array}
[c]{c}%
\begin{array}
[c]{c}%
S_{1}%
\end{array}
\\
\\
\\%
\begin{array}
[c]{c}%
S_{2}%
\end{array}
\end{array}
\overset{\overset{%
\begin{array}
[c]{c}%
\text{Bob}%
\end{array}
}{%
\begin{array}
[c]{ccccccccccccc}%
S_{1}^{\prime} &  &  &  &  &  &  &  &  &  &  &  & S_{2}^{\prime}%
\end{array}
}}{%
\begin{tabular}
[c]{c|c}%
$\underset{}{%
\begin{array}
[c]{ccc}%
(1,1) &  & (1,1)\\
(-3,1) &  & (1,1)
\end{array}
}$ & $\underset{}{%
\begin{array}
[c]{ccc}%
(1,1) &  & (-3,1)\\
(1,1) &  & (1,1)
\end{array}
}$\\\hline
$\overset{}{%
\begin{array}
[c]{cc}%
(-3,1) & (1,1)\\
(1,1) & (1,1)
\end{array}
}$ & $\overset{}{%
\begin{array}
[c]{cc}%
(5,1) & (1,1)\\
(1,1) & (1,1)
\end{array}
}$%
\end{tabular}
\ \ \ }, \label{Game1}%
\end{equation}
for which we consider the strategy pair $(p^{\star},q^{\star})=(1,1/2)$. We
use Eqs.~(\ref{Collect1}) under the assumption $0\leq\Delta$, where $\Delta$
is defined by Eq.~(\ref{delta}), to obtain Nash inequalities for the game
(\ref{Game1}) as%

\begin{gather}
\Pi_{A}(1,1/2)-\Pi_{A}(p,1/2)=-(1/2)(1-p)\left[  2-\Delta\right]
\geq0,\nonumber\\
\Pi_{B}(1,1/2)-\Pi_{A}(1,q)=0.
\end{gather}
As $0\leq(1-p)\leq1$, for this game, the strategy pair $(p^{\star},q^{\star
})=(1,1/2)$ exists as a NE when $2<\Delta$. The converse is also true in that
when $2<\Delta$ the strategy pair $(p^{\star},q^{\star})=(1,1/2)$ becomes a
NE. That is, for the considered game and the strategy pair, Nash and Bell's
inequalities become equivalent.%

\begin{equation}%
\begin{array}
[c]{c}%
\text{Alice}%
\end{array}%
\begin{array}
[c]{c}%
\begin{array}
[c]{c}%
S_{1}%
\end{array}
\\
\\
\\%
\begin{array}
[c]{c}%
S_{2}%
\end{array}
\end{array}
\overset{\overset{%
\begin{array}
[c]{c}%
\text{Bob}%
\end{array}
}{%
\begin{array}
[c]{ccccccccccccc}%
S_{1}^{\prime} &  &  &  &  &  &  &  &  &  &  &  & S_{2}^{\prime}%
\end{array}
}}{%
\begin{tabular}
[c]{c|c}%
$\underset{}{%
\begin{array}
[c]{ccc}%
(1,1) &  & (-3,1)\\
(9,1) &  & (1,1)
\end{array}
}$ & $\underset{}{%
\begin{array}
[c]{ccc}%
(1,1) &  & (9,1)\\
(-3,1) &  & (1,1)
\end{array}
}$\\\hline
$\overset{}{%
\begin{array}
[c]{cc}%
(13,1) & (5,1)\\
(5,1) & (1,1)
\end{array}
}$ & $\overset{}{%
\begin{array}
[c]{cc}%
(-7,1) & (1,1)\\
(1,1) & (5,1)
\end{array}
}$%
\end{tabular}
\ \ }. \label{Game2}%
\end{equation}
Similarly, now considering the game (\ref{Game2}) for the same strategy pair,
we obtain Nash inequalities for the strategy pair $(p^{\star},q^{\star
})=(1,1/2)$ as follows and with the assumption that $\Delta<0$,%

\begin{gather}
\Pi_{A}(1,1/2)-\Pi_{A}(p,1/2)=-(1/2)(1-p)\left[  2+\Delta\right]
\geq0,\nonumber\\
\Pi_{B}(1,1/2)-\Pi_{A}(1,q)=0.
\end{gather}
For this game, the strategy pair $(p^{\star},q^{\star})=(1,1/2)$ exists as a
NE when $\Delta<-2$. The converse is also true in that when $\Delta<-2$ the
strategy pair $(p^{\star},q^{\star})=(1,1/2)$ becomes a NE. That is, for the
considered game and the strategy pair, Nash and Bell's inequalities becomes
equivalent. The strategy pair $(p^{\star},q^{\star})=(1,1/2)$ therefore exists
as a NE in both the games (\ref{Game1}, \ref{Game2}) when Bell's inequality is violated.

As (\ref{Game1}, \ref{Game2}) are especially-designed games, their classical
versions do not have an existing name. The behavior of these games changes
from their quantum pay-off versions in that the particular mixed strategy pair
$(p^{\star},q^{\star})=(1,1/2)$ can exist only in the quantum versions of
these games.

\section{Discussion}

In the quantum games considered in this paper, the players' strategies are
classical consisting of convex linear combinations---with real
coefficients---of their pure classical strategies, whereas the players'
payoffs are obtained directly from the set of quantum probabilities that
underlie the playing of the game. We consider the probabilistic form of Bell's
inequality that can be violated by the set of quantum probabilities. We then
show that there exist such games in which a classical pair of strategies can
be a NE only when the underlying probabilities of the game are truly quantum
mechanical in that they violate Bell's inequality. In the usual approach to
quantum games, a game is given, or known, and pairs of quantum strategies,
consisting of unitary transformations, are determined that constitute a NE. In
a striking contrast this usual approach, a classical strategy pair is
considered as given whereas the set of games is then determine for which that
classical strategy pair becomes a NE only when Bell's inequality is violated.

Some of the earliest criticisms \cite{EnkPike} of quantum games questioned
whether such games are genuinely quantum mechanical. It was suggested that the
violation of Bell's inequality can decidedly determine whether a quantum game
is genuinely quantum or not. Although deriving Bell's inequality does not
require quantum theory, its violation is a well established feature that is
achieveable only in the truely quantum mechanical regime.

As the players' strategies even in the quantum game are restricted to the
classical ones, and the players' payoff relations are obtained from the set of
underlying quantum mechanical probabilities, our approach is not susceptible
to the Enk and Pike type argument \cite{EnkPike}---stating that a quantum game
with quantum mechanical strategies can be considered equivalent to another
extended classical game. In our approach this criticism is avoided as the
players' strategies in both the classical and quantum games remain identical.

The generalized EPR setting used in this paper assumes that the repeated runs
of the EPR experiment are performed in order to obtain the expected values of
quantum mechanical observables. In particular, it is not the case that the
measurement outcomes of an individual run lead to the players revising their
strategic moves in the next run in view of their payoffs in the previous run,
as is the case in repeated games. A study of repeated quantum games using
generalized EPR experiments is an open question for future work.

Note that although the players' strategies are classical, quantum mechanics is
central to the setting of the considered quantum game. Players' payoff
relations have underlying quantum probability distributions. The physical
system that is used to play this game is the standard EPR type apparatus
involving Stern-Gerlach type measurements. Local unitary transformations are
used as the players' strategies in the standard schemes to play quantum games
whereas classical strategies, akin to rotating the arms of an EPR apparatus,
are the players' strategies in this present paper \cite{IqbalWeigert,
Iqbalepr1,Iqbalepr2,Iqbalepr3,Iqbalepr4,Iqbalepr5,Iqbalepr6,Iqbalepr7,Iqbalepr8}%
. We identify sets of games in which, for a considered classical mixed
strategy, the Nash inequality becomes equivalent to Bell's inequality.

The results of this paper can be extended to multi-player games. This would
involve consideration of the N-partite Bell's inequality \cite{Wu et al}---a
situation in which use of geometric algebra has been shown to offer a
tractable setting for the analysis of N-partite interactions \cite{Iqbalepr8}.
Also, consideration of two-player games with multi strategies will involve
Bell's inequalities with many observables \cite{Ryu et al}.

\section*{Appendix A}

When translated in terms of the probability set $\epsilon_{j},$ the locality
requirements can be expressed as%

\begin{equation}%
\begin{array}
[c]{cc}%
\epsilon_{1}+\epsilon_{2}=\epsilon_{5}+\epsilon_{6}, & \epsilon_{1}%
+\epsilon_{3}=\epsilon_{9}+\epsilon_{11},\\
\epsilon_{9}+\epsilon_{10}=\epsilon_{13}+\epsilon_{14}, & \epsilon
_{5}+\epsilon_{7}=\epsilon_{13}+\epsilon_{15},\\
\epsilon_{3}+\epsilon_{4}=\epsilon_{7}+\epsilon_{8}, & \epsilon_{11}%
+\epsilon_{12}=\epsilon_{15}+\epsilon_{16},\\
\epsilon_{2}+\epsilon_{4}=\epsilon_{10}+\epsilon_{12}, & \epsilon_{6}%
+\epsilon_{8}=\epsilon_{14}+\epsilon_{16}.
\end{array}
\label{LocalityConstraint}%
\end{equation}
Cereceda \cite{Cereceda} reports a convenient solution of the system
(\ref{Normalization}, \ref{LocalityConstraint}) for which the set of
probabilities $\upsilon=\left\{  \epsilon_{2},\text{ }\epsilon_{3},\text{
}\epsilon_{6},\text{ }\epsilon_{7},\text{ }\epsilon_{10},\text{ }\epsilon
_{11},\text{ }\epsilon_{13},\text{ }\epsilon_{16}\right\}  $ is expressed in
terms of the remaining set of probabilities i.e.%

\begin{equation}%
\begin{array}
[c]{l}%
\epsilon_{2}=(1-\epsilon_{1}-\epsilon_{4}+\epsilon_{5}-\epsilon_{8}%
-\epsilon_{9}+\epsilon_{12}+\epsilon_{14}-\epsilon_{15})/{\small 2},\\
\epsilon_{3}=(1-\epsilon_{1}-\epsilon_{4}-\epsilon_{5}+\epsilon_{8}%
+\epsilon_{9}-\epsilon_{12}-\epsilon_{14}+\epsilon_{15})/{\small 2},\\
\epsilon_{6}=(1+\epsilon_{1}-\epsilon_{4}-\epsilon_{5}-\epsilon_{8}%
-\epsilon_{9}+\epsilon_{12}+\epsilon_{14}-\epsilon_{15})/{\small 2},\\
\epsilon_{7}=(1-\epsilon_{1}+\epsilon_{4}-\epsilon_{5}-\epsilon_{8}%
+\epsilon_{9}-\epsilon_{12}-\epsilon_{14}+\epsilon_{15})/{\small 2},\\
\epsilon_{10}=(1-\epsilon_{1}+\epsilon_{4}+\epsilon_{5}-\epsilon_{8}%
-\epsilon_{9}-\epsilon_{12}+\epsilon_{14}-\epsilon_{15})/{\small 2},\\
\epsilon_{11}=(1+\epsilon_{1}-\epsilon_{4}-\epsilon_{5}+\epsilon_{8}%
-\epsilon_{9}-\epsilon_{12}-\epsilon_{14}+\epsilon_{15})/{\small 2},\\
\epsilon_{13}=(1-\epsilon_{1}+\epsilon_{4}+\epsilon_{5}-\epsilon_{8}%
+\epsilon_{9}-\epsilon_{12}-\epsilon_{14}-\epsilon_{15})/{\small 2},\\
\epsilon_{16}=(1+\epsilon_{1}-\epsilon_{4}-\epsilon_{5}+\epsilon_{8}%
-\epsilon_{9}+\epsilon_{12}-\epsilon_{14}-\epsilon_{15})/{\small 2}.
\end{array}
\end{equation}

\end{document}